\documentclass[a4paper]{article}

\usepackage{icrc2013}
\usepackage[english]{babel}
\title{Monte Carlo comparison of medium-size telescope designs for the Cherenkov Telescope Array}

\shorttitle{ICRC 2013 Template}

\authors{
T. Jogler$^{1}$,
M. D. Wood$^{1}$,
J. Dumm$^{2}$,
A. Bouvier$^{3}$,
for the CTA Consortium.
}

\afiliations{
$^1$ SLAC National Accelerator Laboratory, 2575 Sand Hill Road M/S 29,
Menlo Park, CA 94025, USA \\
$^2$ University of Minnesota\\
$^3$ University of California Santa Cruz\\
}

\email{jogler@slac.stanford.edu}

\abstract{The Cherenkov Telescope Array (CTA) is a future very high
  energy gamma-ray observatory. CTA will be comprised of small-,
  medium- and large-size telescopes covering an energy range from tens
  of GeV to hundreds of TeV and will surpass existing telescopes in
  sensitivity by an order of magnitude. The aim of our study is to
  find the optimal design for the medium-size telescopes (MSTs), which
  will determine the sensitivity in the key energy range between a few
  hundred GeV to about ten TeV.  To study the effect of the telescope
  design parameters on the array performance, we simulated arrays of
  61 MSTs with 120~m spacing and a variety of telescope
  configurations.  We investigated the influence of the primary
  telescope characteristics including optical resolution, pixel size,
  and light collection area on the total array performance with a
  particular emphasis on telescope configurations with imaging
  performance similar to the proposed Davies-Cotton (DC) and
  Schwarzschild-Couder (SC) MST designs.  We compare the performance
  of these telescope designs, especially the achieved gamma-ray
  angular resolution and differential point-source
  sensitivity. Finally we investigate the performance of different
  array sizes to demonstrate impacts of financial constraints on the
  number of telescopes.}

\keywords{CTA, gamma-rays, monte carlo, simulations}

\begin{document}
\maketitle

\section{Introduction}
The Cherenkov Telescope Array (CTA) is the future next generation
Imaging Atmospheric Cherenkov Telescope (IACT) observatory. CTA aims
to surpass the current IACT systems like HESS, MAGIC and VERITAS by an
order of magnitude in sensitivity and enlarge the observable energy
range from a few tens of GeV to far beyond one hundred
TeV~\cite{whitepaper}. To achieve this broad energy range and high
sensitivity, CTA will be comprised of three different telescope
sizes. These are denoted according to their mirror diameter into
large-size telescopes, medium-size telescopes (MSTs), and small-size
telescopes.

In this paper we investigate the effect of the optical point-spread
function and the camera pixel size on the achievable point-source
sensitivity. We investigate MSTs since they are most sensitive in the
energy range where the best angular resolution is achieved and small
pixels sizes are most feasible.

The main motivation for this study is to determine if the current,
well-tested single-mirror design (Davies-Cotton, DC) or a new
two-mirror design (Schwarzschild-Couder, SC) would be the best choice
for the medium-size CTA telescopes.  
The SC telescopes can achieve much smaller optical point-spread
function (PSF), minimal aberrations over a wider field of view, and a
smaller plate scale, allowing for a more compact camera.
However, SC telescopes require many more readout channels and more
complicated mirror designs that increase their price compared to a DC
telescope with similar mirror area. We simulate idealized telescope
parameters for both designs and compare their gamma-ray PSF and
point-source sensitivity.  We also investigate the sensitivity of
arrays with different numbers of telescopes with a specific focus on
the performance of arrays with and without a US contribution of 36
telescopes.
  
\section{Simulations}

\begin{figure}[h]
  \includegraphics[width=0.5\textwidth]{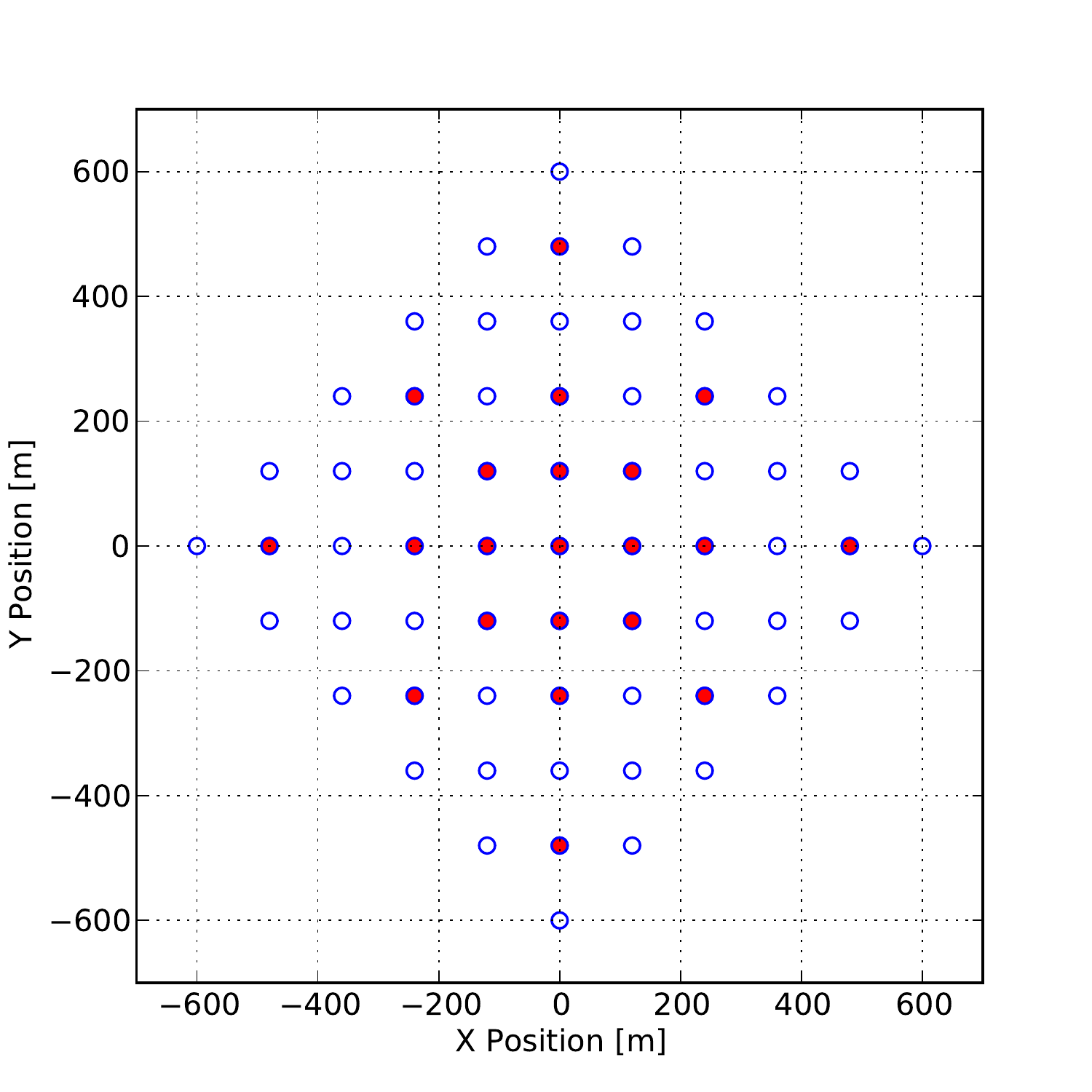}
  \caption{\label{FIG:LAYOUT} Comparison of the telescope positions of
    the two simulated array geometries with 61 telescopes (open blue
    circles) and 21 telescopes (filled red circles).}
\end{figure}

Gamma-ray and proton air showers were simulated with the CORSIKA Monte
Carlo (MC) package \cite{heck1998} and the QGSJet-II hadronic
interaction model \cite{2006NuPhS.151..143O}.  Simulations were
performed for an array at an elevation of 2000~m and geomagnetic field
configuration similar to the proposed southern hemisphere sites.
Showers were simulated at 20$^\circ$ zenith angle over the energy
range from 10~GeV to 30~TeV.  All simulations use the same array
layout comprising 61 telescopes forming a square with 120~m
inter-telescope spacing. Each array is composed of identical
telescopes.  To study the performance of a reduced array without a US
contribution we additionally simulated a 21 telescope array by
using a subset of telescopes from the 61 telescope layout (see
Figure \ref{FIG:LAYOUT}).

\subsection{Telescope Designs}
We simulated a range of optical PSFs and pixel sizes that bracket the
imaging performance of the DC- and SC-like telescope designs.  The
proposed designs for the DC- and SC-MST have a 68\% optical PSF
containment radius (R$_{68}$) of 0.04$^\circ$-0.1$^\circ$ over the FoV
and 0.02$^\circ$-0.04$^\circ$ over the FoV, respectively. The
simulated pixel sizes (R$_\textrm{pix}$) are between 0.06$^\circ$ and
0.16$^\circ$.  We use the configurations with
R$_{68}$/R$_\textrm{pix}$ of 0.08$^\circ$/0.16$^\circ$ and
0.02$^\circ$/0.06$^\circ$ as representative of configurations with DC-
and SC-like imaging performance, respectively.  For both
configurations we assume a field of view of $8^{\circ}$.

Besides imaging resolution, the other important characteristic of the
telescope optical system is the total effective light collection area, 
\begin{equation}
  A_\textrm{opt}(\lambda_{0},\lambda_{1}) = 
  A_{M}\int_{\lambda_{0}}^{\lambda_{1}}P(\lambda)
  \epsilon(\lambda)d\lambda,
\end{equation}
which is the product of the mirror area ($A_{M}$) with the weighted
average of the optical efficiency ($\epsilon(\lambda)$) with a
Cherenkov-like spectral distribution ($P(\lambda)$).  We compute the
effective light collection area over the wavelength interval of 250 nm
to 700 nm and a Cherenkov spectral distribution calculated for an
emission height of 10~km.
We simulated two telescope configurations chosen to be representative
of DC- and SC-MST designs of equal cost.  Both telescopes have an
aperture of 10~m and $A_\textrm{opt}$ of 6.3~m$^{2}$ (SC-MST) and
14.9~m$^{2}$ (DC-MST).

\subsection{Trigger and DAQ}
A simplified detector model is used to simulate the time-integrated
signal in each pixel.  The pixel signal is the sum of the detected
Cherenkov photo electrons (phes) and a noise component modeled as the
sum of a Poisson-distributed NSB term and a Gaussian-distributed
electronics noise term with an RMS of 0.1~phe per channel.  The mean
NSB amplitude in each pixel is $\Delta\Omega (A_\textrm{opt}/11.8$
m$^{2})\times (100$ phe deg$^{-2})$ where $\Delta\Omega$ is the pixel
solid angle.  The NSB amplitude was chosen to be representative of the
sky brightness of an extragalactic observation field and an
integration gate of 10~ns.  Each telescope camera containing more than
60~phe is assumed to trigger, and at least two telescopes must trigger
to produce an array trigger. All array triggered events are further
processed.

\subsection{Analysis}
Reconstruction of the telescope image data into event-level parameters
proceeds in three stages.  First, an image cleaning is performed to
select pixels with statistically significant signal amplitude.  The
shower trajectory is then reconstructed using a geometric analysis of
the moments of the light distribution in each camera.  Finally, a
likelihood-based reconstruction is performed using templates for the
light distribution in each telescope derived from MC simulations.  In
addition to the event trajectory and energy, a number of parameters
useful for gamma-hadron discrimination are calculated such as the
goodness-of-fit of the telescope images with respect to the image
templates.  Background suppression is performed with the TMVA boosted
decision tree (BDT) method \cite{tmva}.  The decision trees (DTs) are
trained with independent samples of simulated gamma-ray and proton
events.  Energy-dependent cuts on the BDT output variable and
$\theta^2$, the squared angular separation between the reconstructed
and source directions, are optimized under the assumption of a
point-like source distribution with an intensity equal to 1\% of the
Crab Nebula flux.

\subsection{Telescope array layouts}
The price for an individual telescope and the total budget of CTA are
not yet finalized, so we investigated possible array configurations
with reduced number of telescopes. The array layouts comprise of 21,
25, 41 and 61 Telescopes. These studies demonstrate that the number of
telescopes has a significant impact on the total sensitivity
regardless of the telescope type. The number of telescopes are chosen
to match the MST number in the CTA configuration I and
E~\cite{Bernlohr2012} and the extension of I and E by up to 36 US
telescopes. In the case of the configuration I we replaced the 4
Large-size telescopes with medium-size telescopes. The final CTA
layout will have a much better sensitivity at low Energies
($E<100\textrm{ GeV}$) and above 3 TeV due to the large- and
small-size telescopes.

\section{Results \& Conclusions}
\begin{figure}[ht!]
\includegraphics[width=0.5\textwidth]{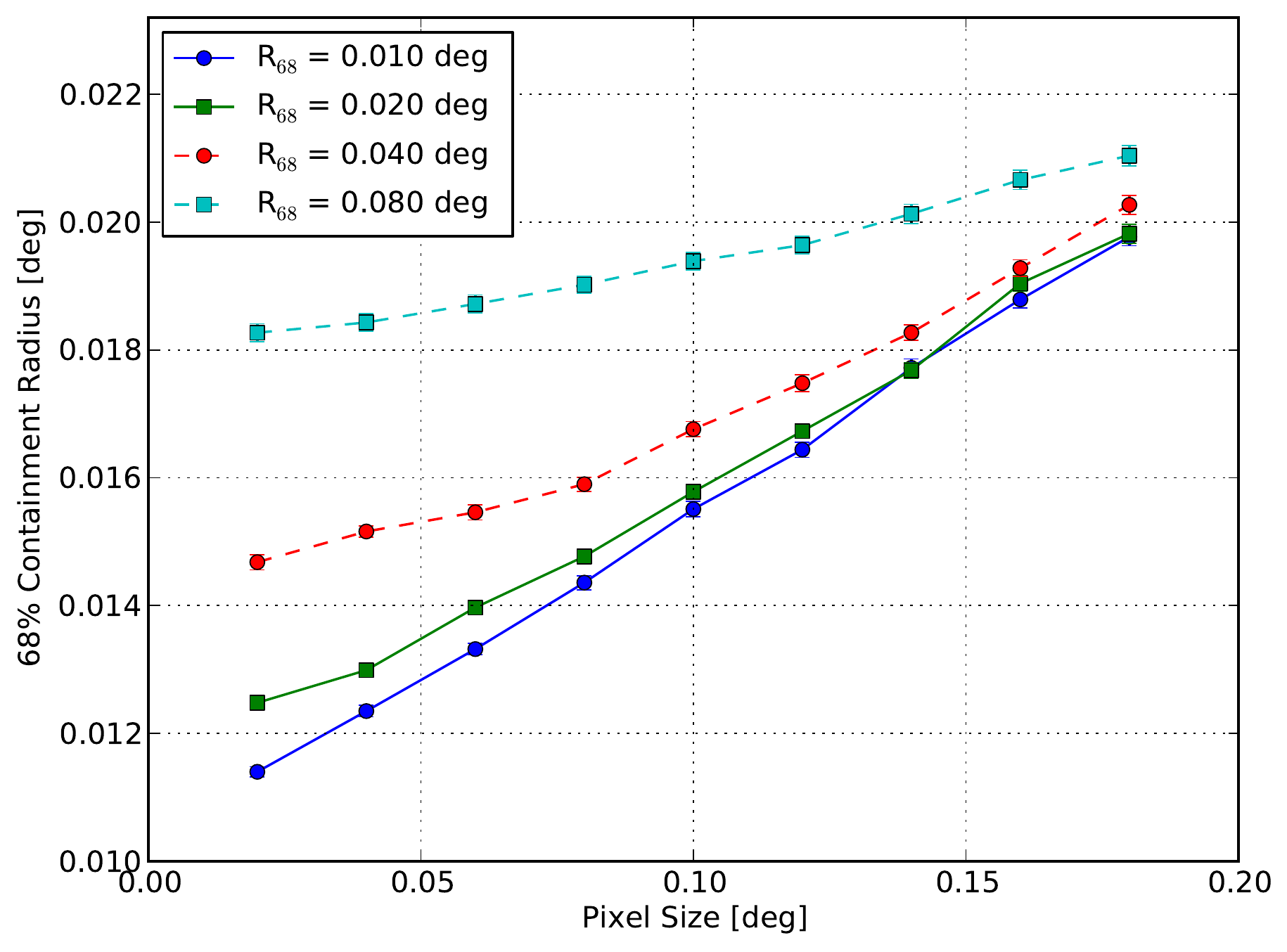}
\includegraphics[width=0.5\textwidth]{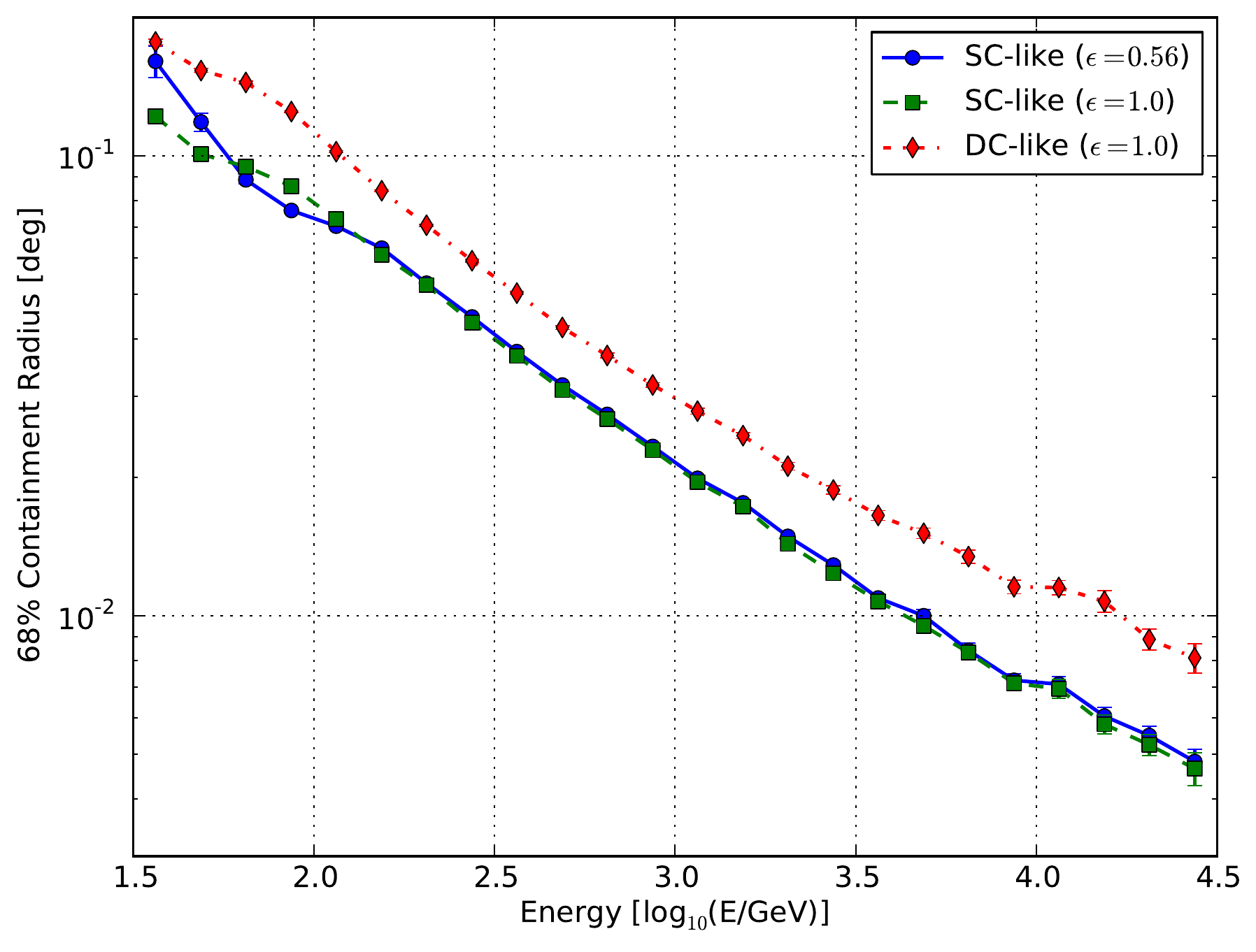}
\caption{\label{FIG:PSF}\textbf{Top:} 68\% containment radius of the gamma-ray PSF
    at 1~TeV versus pixel size shown for a 61 telescope MST array
    composed of telescopes with increasing 68\% optical PSF
    containment radii: 0.01$^\circ$ (blue circles and solid line),
    0.02$^\circ$ (green squares with solid line), 0.04$^\circ$ (red
    circles and dashed line), 0.08$^\circ$ (cyan squares and dashed
    line).  \textbf{Bottom:} 68\% containment radius of the gamma-ray
    PSF versus gamma-ray energy for an array composed of telescopes
    with SC-like imaging performance with effective light collection
    area scaled by 0.56 (blue circles and solid line) and 1.0 (green
    squares and dashed line) and a DC-like imaging performance (red
    diamonds and dot-dashed line). }
\end{figure}

\begin{figure}[ht!]
  \includegraphics[width=0.5\textwidth]{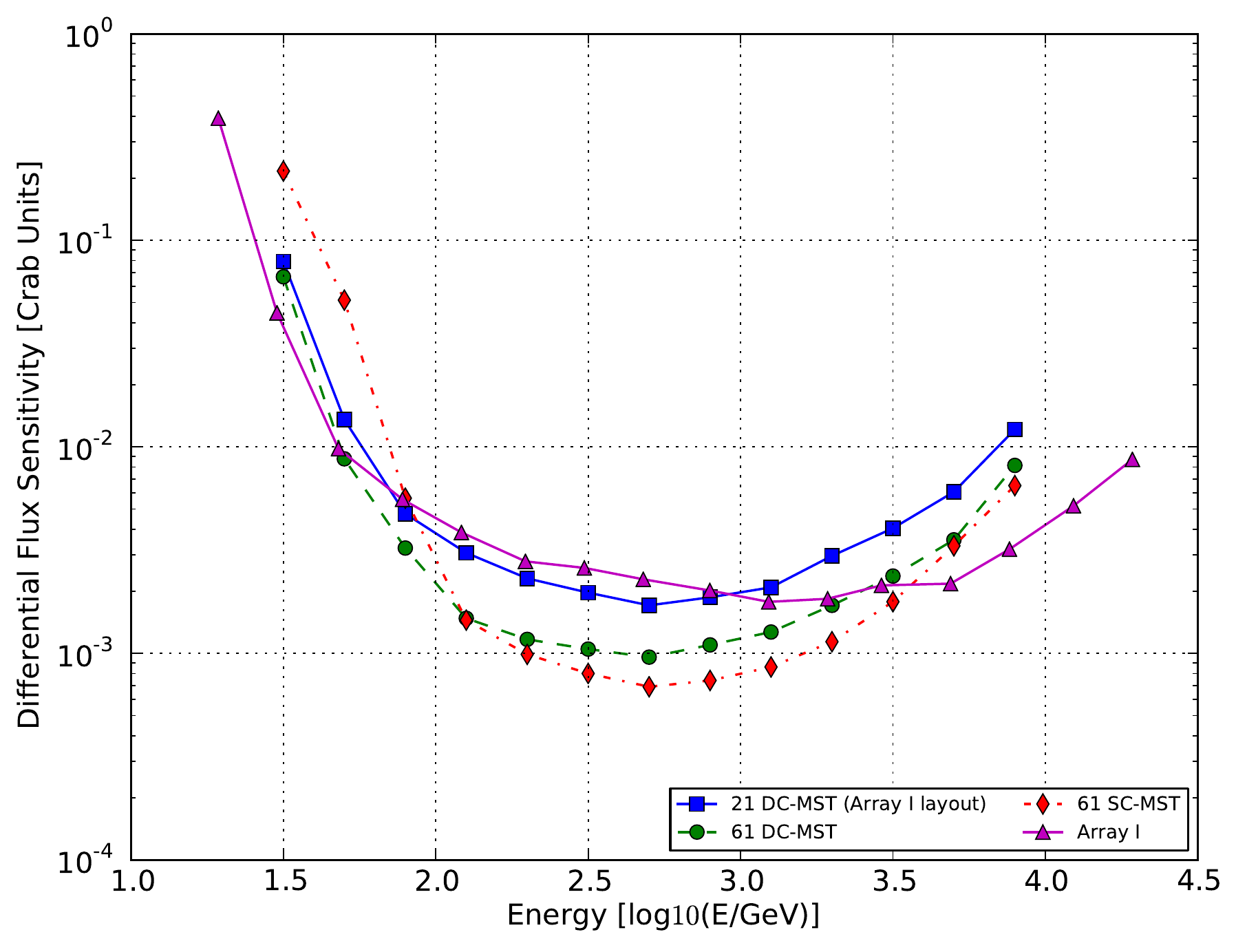}
  \includegraphics[width=0.5\textwidth]{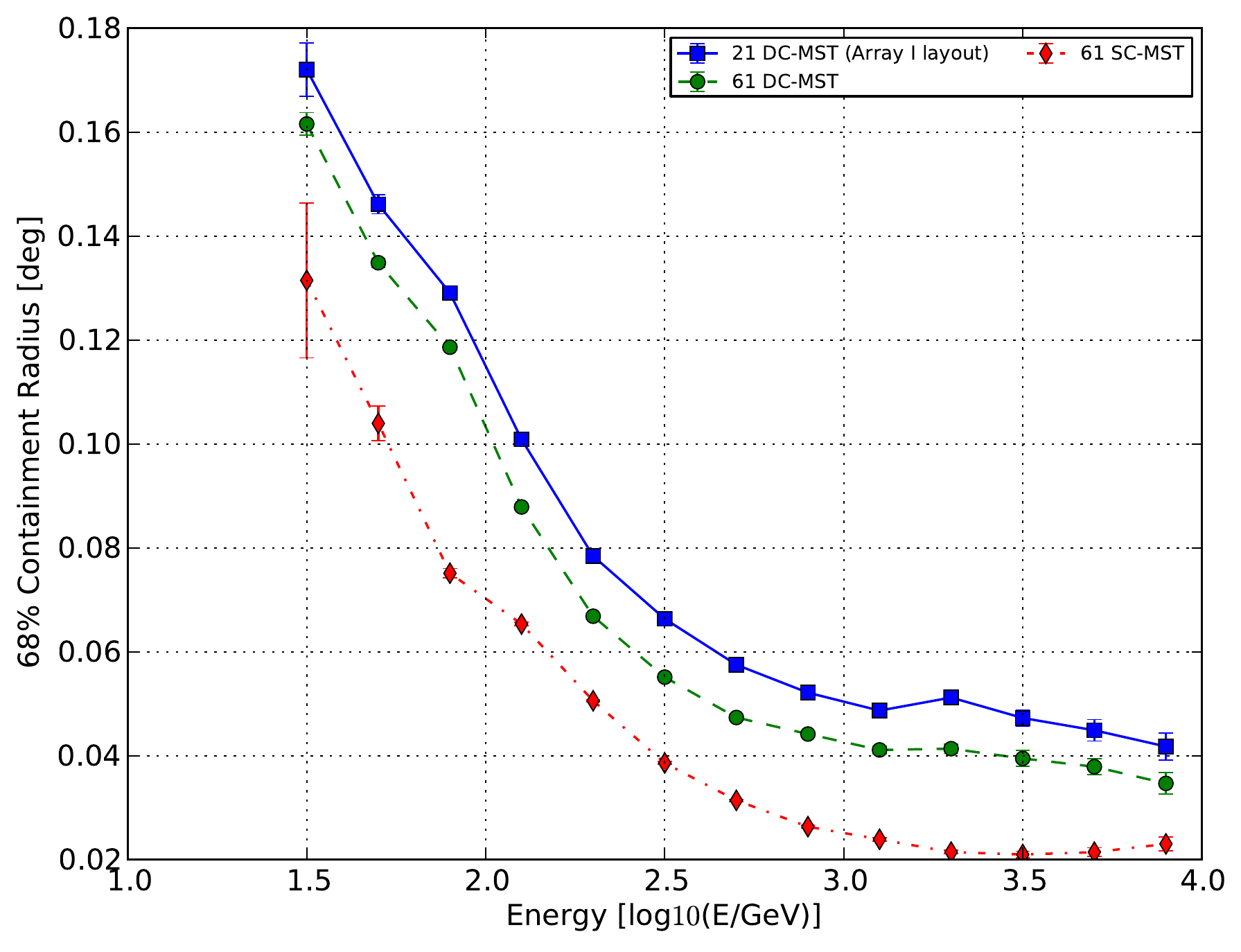}
  \caption{\label{FIG:SENSITIVITY}\textbf{Top:} Differential
    point-source sensitivity for an array of 21 DC-MSTs (blue squares
    and solid line), 61 DC-MSTs (green circles and dashed line), 61
    SC-MSTs (red diamonds and dot dashed line), and one of the
    proposed CTA designs (array I) with 18 DC-like MSTs
    \cite{Bernlohr2012} (magenta triangles and solid line). The array
    I comprises large-size and small-size telescopes not included in
    our simulation. \textbf{Bottom:} 68\% containment radius of the
    gamma-ray PSF versus gamma-ray energy for the three array
    configurations shown in the left figure.}

\end{figure}

The gamma-ray PSF improves when reducing the pixel size as long as the
optical PSF is smaller than the pixel size.  The SC-like telescope
array shows a 40\% improved gamma-ray PSF compared to the DC-like
telescopes at all energies (see Fig.~\ref{FIG:PSF}).
Figure \ref{FIG:SENSITIVITY} shows that the SC-like array has a
30--40\% better differential point-source sensitivity relative to the
DC-like array at energies above 100~GeV which is mainly due to the
improved gamma-ray PSF.  Below 100~GeV the smaller light collection
area of the SC-like telescope configuration is a disadvantage
resulting in a higher reconstruction energy threshold and an equal or
slightly worse differential sensitivity.

While the SC-like array is more sensitive compared to the DC-like
array, no SC telescope has been built to date. The results presented
here provide encouragement to build an SC prototype telescope to test
if the performance can be achieved under realistic conditions.
However, our studies show that the extension of the array by the US
contribution will improve the sensitivity of CTA in its key energy
range (0.3 to 3 TeV) by a factor of 2--3 depending on the type of
telescopes choosen. That would reflect in an decrease of observation
time by a factor 4--9 and would allow the study of many more objects
or a much faster survey of the gamma-ray sky.


\section{Acknowledgments}
We gratefully acknowledge support from the agencies and organizations 
listed in this page: \url{http://www.cta-observatory.org/?q=node/22}



\bibliographystyle{aipproc}   

\bibliography{sample}

\IfFileExists{\jobname.bbl}{}
 {\typeout{}
  \typeout{******************************************}
  \typeout{** Please run "bibtex \jobname" to optain}
  \typeout{** the bibliography and then re-run LaTeX}
  \typeout{** twice to fix the references!}
  \typeout{******************************************}
  \typeout{}
 }

\end{document}